\documentclass[superscriptaddress,preprintnumbers,amsmath,amssymb,prd,nofootinbib,preprint]{revtex4-1}
\pdfoutput=1
\usepackage{graphicx}
\usepackage{epstopdf}
\usepackage{dcolumn}
\usepackage{bm}
\usepackage{hyperref}
\usepackage{color}
\usepackage{amsmath}
\usepackage{cancel}
\usepackage{xpatch}

\makeatletter
\def\l@subsubsection#1#2{}
\patchcmd{\@ssect@ltx}
    {\addcontentsline{toc}{#1}{\protect\numberline{}#8}}
    {}
    {}
    {}
\makeatother
\begin{document}

\def\a{\alpha}
\def\b{\beta}
\def\c{\varepsilon}
\def\d{\delta}
\def\e{\epsilon}
\def\f{\phi}
\def\g{\gamma}
\def\h{\theta}
\def\k{\kappa}
\def\l{\lambda}
\def\m{\mu}
\def\n{\nu}
\def\p{\psi}
\def\q{\partial}
\def\r{\rho}
\def\s{\sigma}
\def\t{\tau}
\def\u{\upsilon}
\def\v{\varphi}
\def\w{\omega}
\def\x{\xi}
\def\y{\eta}
\def\z{\zeta}
\def\D{\Delta}
\def\G{\Gamma}
\def\H{\Theta}
\def\L{\Lambda}
\def\F{\Phi}
\def\P{\Psi}
\def\S{\Sigma}

\def\o{\over}
\def\beq{\begin{align}}
\def\eeq{\end{align}}
\newcommand{\gsim}{ \mathop{}_{\textstyle \sim}^{\textstyle >} }
\newcommand{\lsim}{ \mathop{}_{\textstyle \sim}^{\textstyle <} }
\newcommand{\vev}[1]{ \left\langle {#1} \right\rangle }
\newcommand{\bra}[1]{ \langle {#1} | }
\newcommand{\ket}[1]{ | {#1} \rangle }
\newcommand{\EV}{ {\rm eV} }
\newcommand{\KEV}{ {\rm keV} }
\newcommand{\MEV}{ {\rm MeV} }
\newcommand{\GEV}{ {\rm GeV} }
\newcommand{\TEV}{ {\rm TeV} }
\newcommand{\1}{\mbox{1}\hspace{-0.25em}\mbox{l}}
\newcommand{\headline}[1]{\noindent{\bf #1}}
\def\diag{\mathop{\rm diag}\nolimits}
\def\Spin{\mathop{\rm Spin}}
\def\SO{\mathop{\rm SO}}
\def\O{\mathop{\rm O}}
\def\SU{\mathop{\rm SU}}
\def\U{\mathop{\rm U}}
\def\Sp{\mathop{\rm Sp}}
\def\SL{\mathop{\rm SL}}
\def\tr{\mathop{\rm tr}}
\def\mpl{M_{\rm Pl}}

\def\IJMP{Int.~J.~Mod.~Phys. }
\def\MPL{Mod.~Phys.~Lett. }
\def\NP{Nucl.~Phys. }
\def\PL{Phys.~Lett. }
\def\PR{Phys.~Rev. }
\def\PRL{Phys.~Rev.~Lett. }
\def\PTP{Prog.~Theor.~Phys. }
\def\ZP{Z.~Phys. }

\def\dd{\mathrm{d}}
\def\ff{\mathrm{f}}
\def\BH{{\rm BH}}
\def\inf{{\rm inf}}
\def\ev{{\rm evap}}
\def\eq{{\rm eq}}
\def\SM{{\rm sm}}
\def\Mpl{M_{\rm Pl}}
\def\GeV{{\rm GeV}}
\newcommand{\Red}[1]{\textcolor{red}{#1}}
\newcommand{\TL}[1]{\textcolor{blue}{\bf TL: #1}}


\begin{center}
\vspace{1cm}
{\Large \bf The paradox of the Standard Model} \\
\vspace{2cm}
{\large Riccardo Barbieri} \\

{\it \small Scuola Normale Superiore, Piazza dei Cavalieri 7, 56126 Pisa, Italy and INFN, Pisa, Italy} \\
\vspace{1cm}
\end{center}
As a way to recall and honour Guido Altarelli I contribute to the discussion on the prospect for the future of Particle Physics. I do this because I am  sure that Guido would have liked the discussion, even though I am not equally sure that he would have agreed on all the opinions I express here. \\

\vspace{4cm}
\begin{center}
A contribution to:\\
"From my vast repertoire: the legacy of Guido Altarelli"\\
S. Forte, A. Levy and G. Ridolfi,eds.
\end{center}

%
%
%
%
%
%
%
%

\newpage

 The triumph of reductionism represented by the Standard Model (SM) is evident, as made particularly manifest by its Lagrangian in Fig.~\ref{SML}. In three  lines one describes in details the behaviour of the constituents of matter with the greatest empirical adequacy at all distances between about $10^{-18}$ and $10^{-7}$ meters and with the possibility to be extended much further.  The synthetic character of PP emerges at full strength. Yet there are strong reasons to think that the SM is not a complete theory. This is the paradox alluded to in the title.  From a practical point of view and in a simplified way the contrast can be summarised by considering two apparently alternative directions
of research:
\begin{itemize}
\item Give the SM for granted and "look elsewhere".
\item Keep testing the SM to learn how to complete it.
\end{itemize}
I actually think that this is a false  alternative. The "look elsewhere" attitude is at the heart of fundamental physics and does not need any defence. On the other hand I see reasons of poor understanding and reasons of incompleteness to keep pursuing the second direction as well. In the first part of this discussion I illustrate these reasons in succession.

\section{The case for precision}

\begin{figure}
   \centering
\includegraphics[width=0.84\textwidth]{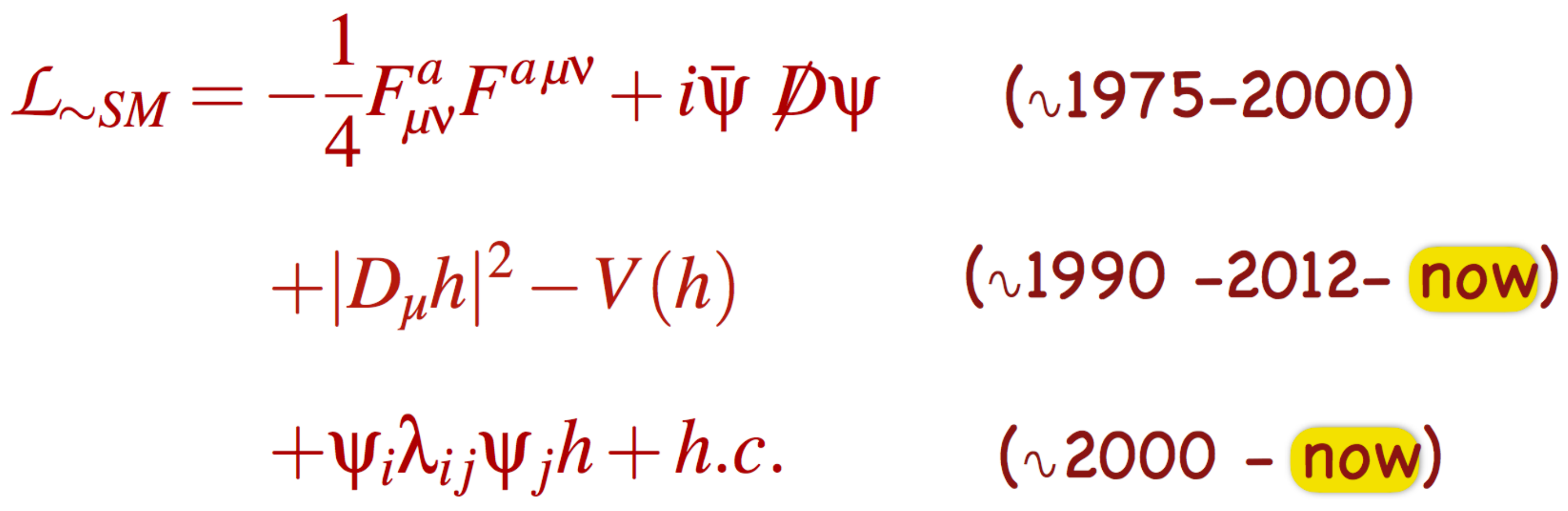}
\caption{The Lagrangian of the SM with the indication, in correspondence to each line, of the approximate dates of their experimental progressive shining.}
  \label{SML}
\end{figure}

Although the three different lines in Fig.~\ref{SML} are no doubt interdependent, I think that the distinction in the dates - so to say - of their respective experimental shining, as indicated in the figure, is totally meaningful.
The tests of the gauge sector of the SM are overwhelming in extension and precision, both in the strong and in the electroweak interactions. The constrained structure of a general gauge Lagrangian plays an important role in this, in contrast to the case of the second and third lines in Fig.~\ref{SML}.

From the experimental side the case is also different. On the second line the tests started indirectly with the electroweak precision measurements and, in a sense, culminated in 2012 with the discovery of the Higgs boson\cite{Aad:2012tfa,Chatrchyan:2012xdj}. Yet the precision at which the Higgs couplings are known at present does not exceed  the $20\%$ level, limited to a few cases\cite{Khachatryan:2016vau}. Precision in PP is a key to understanding. As an example the current measurements of the Higgs couplings allow a compositeness scale for the Higgs boson as low as about 600 GeV, compared to the tens of TeVs in the case, e.g.,  of the electron, whose couplings are much better known. There are reasons to think that the Higgs boson is the least understood particle of the SM. It cannot be the one that it is less precisely measured. 

\begin{figure}
   \centering
\includegraphics[width=0.51\textwidth]{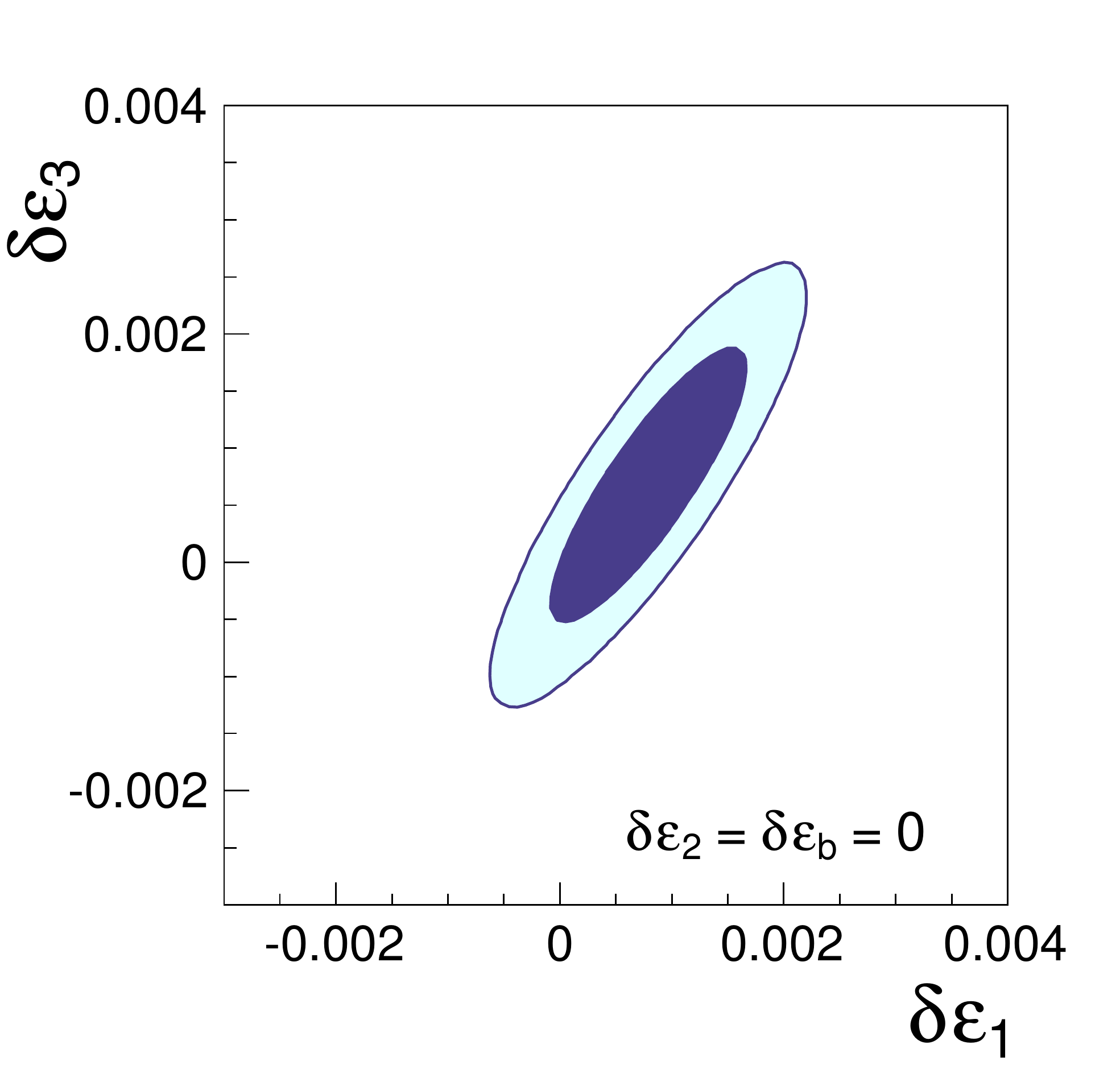}\,
\includegraphics[width=0.47\textwidth]{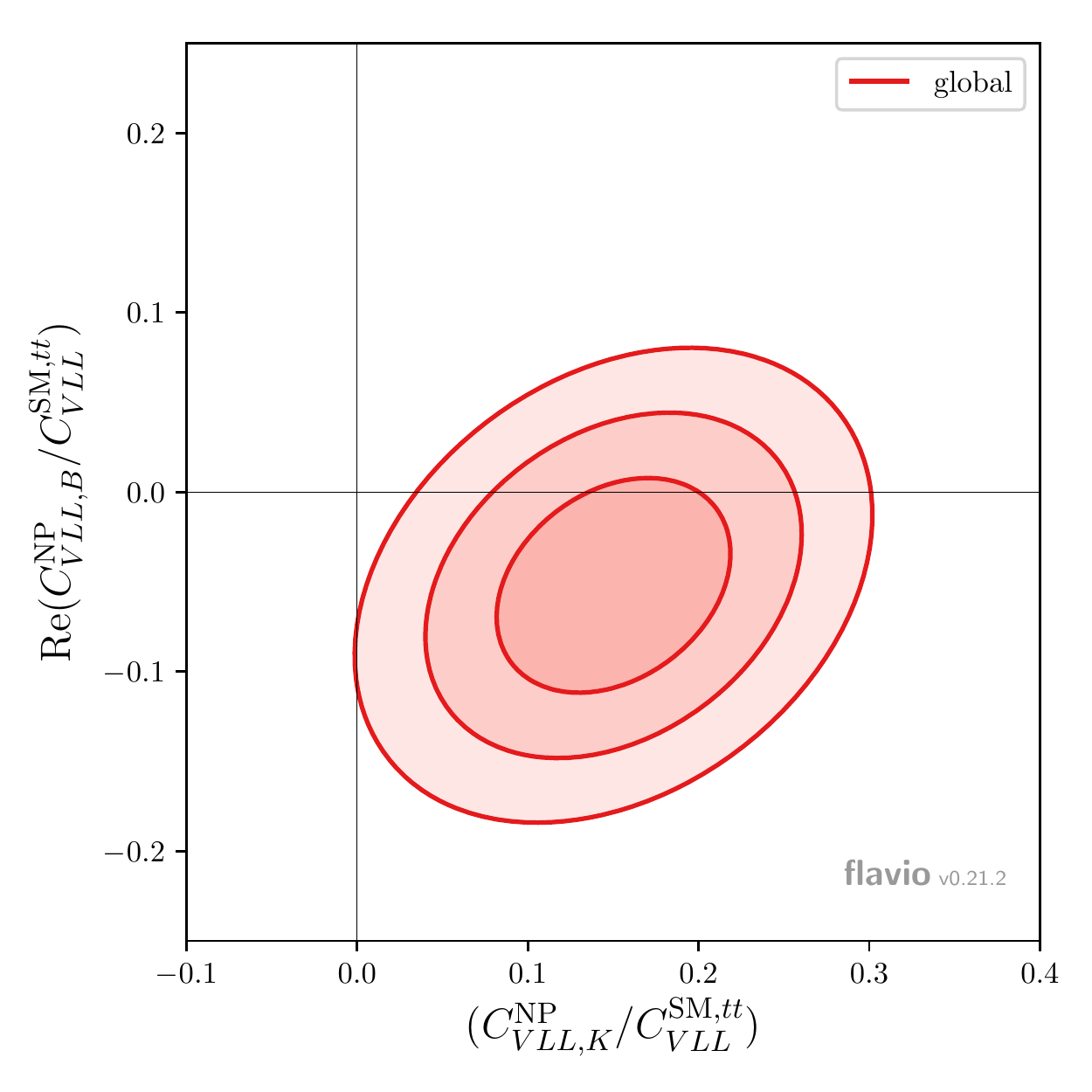}
\caption{LEFT: Deviations of the electroweak precision parameters from the SM values, $\epsilon_1^{SM}=5.21\cdot 10^{-3}$ and $\epsilon_3^{SM}=5.28\cdot 10^{-3}$ (courtesy of the HEPfit group); RIGHT: Relative deviations of the Wilson coefficients of the leading SM operators from the SM  values in $\Delta B_{d,s}=2$, vertical axis, and in $Im(\Delta S=2)$, horizontal axis (courtesy of D. Straub).}
  \label{comp}
\end{figure}

The case of the flavour line, the third one in Fig.~\ref{SML}, is again different. The test of unitarity of the Cabibbo-Kobayashi-Maskawa (CKM) matrix  has strongly progressed in the last twenty  years or so, starting from the measurement of the $\epsilon'/\epsilon$ parameter in K-physics\cite{Fanti:1999nm}. Yet the pure parametric description of the spectrum and mixings of quarks and leptons is a strong drawback  of the SM model.
Admittedly one does not know if further measurements will allow to improve on this situation, but the case for better precision in flavour physics is also very strong. About this, Fig.~\ref{comp} shows an interesting comparison. 
The figure on the left side shows that the genuine electroweak loops in the SM, of about $5\cdot 10^{-3}$ in both $\epsilon_1$ and  $\epsilon_3$\cite{Altarelli:1990zd}, are currently measured at about $20\%$ level, as shown in the figure by their deviations from the SM values\cite{Ciuchini:2014dea}. Interestingly enough the deviations of the Wilson coefficients of the leading SM operators, arising from the dominant Flavour Changing Neutral Currents loops, are bound to be at most of similar $10-20\%$ relative size. This is shown in the figure on the right, where it is assumed that possible New Physics contributions in $\Delta B_s$ and in $\Delta B_d$ scale, relative to each other, like $V_{ts}/V_{td}$ as in the SM. It is well appreciated how important it would be to improve by one order of magnitude the precision on the electroweak loops, as it might be possible in a future $e^+e^-$ facility. The goal of reaching similar relative precision in flavour tests might  be attainable  by an aggressive flavour program that aimed  at collecting $\mathcal{O}(10)$ times the LHCb upgrade luminosity and, in parallel, at significantly improving the efficiency in suitable low $p_T$ events in ATLAS and CMS. 

\section{The incompleteness of the Standard Model}
\label{sec:PC}

\begin{figure}
   \centering
\includegraphics[width=0.90\textwidth]{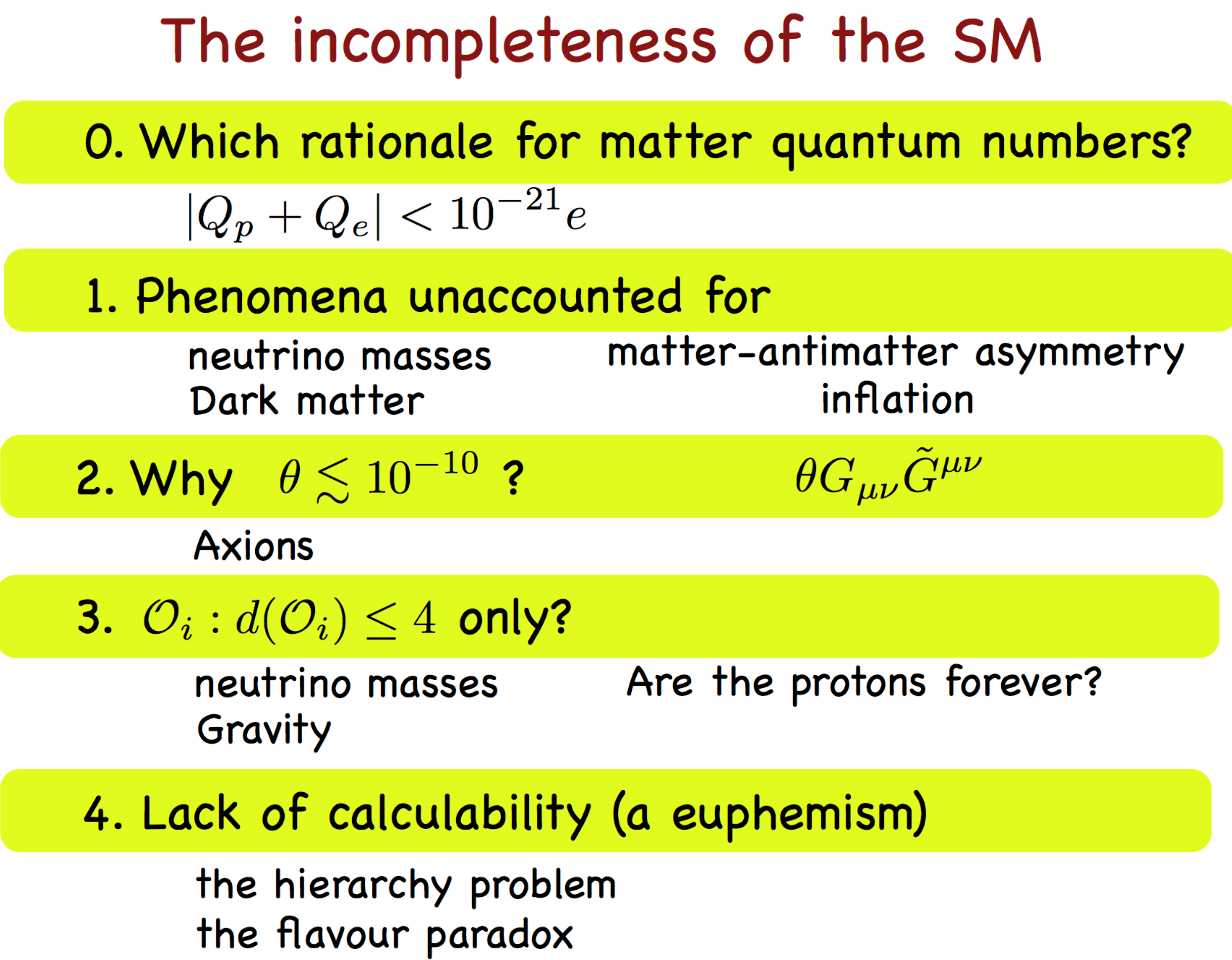}
\caption{A  list of the incompletenesses of the SM}
  \label{SMincomp}
\end{figure}

It is often recalled, rightly so, that the SM does not account for some major phenomena, like neutrino masses, the existence of Dark Matter or the matter-antimatter asymmetry. I find it useful to insert these issues in the broader prospective of seeing the various incompletenesses of the SM all at once, as attempted  in Fig.~\ref{SMincomp}. Most of the entries are well known, to the point that we think that we know the solution for many of them. The trouble is that we do not know if any or almost any of these solutions is right. We lack the corresponding experimental evidence, whose search is at the core of the PP program. For reasons of space here I limit myself to some comments on the hierarchy problem, once again, as due after the (temporary) blank of the LHC on BSM searches.

\subsection{The hierarchy problem, once again}
\label{hierarchy}

\begin{figure}
   \centering
\includegraphics[width=0.80\textwidth]{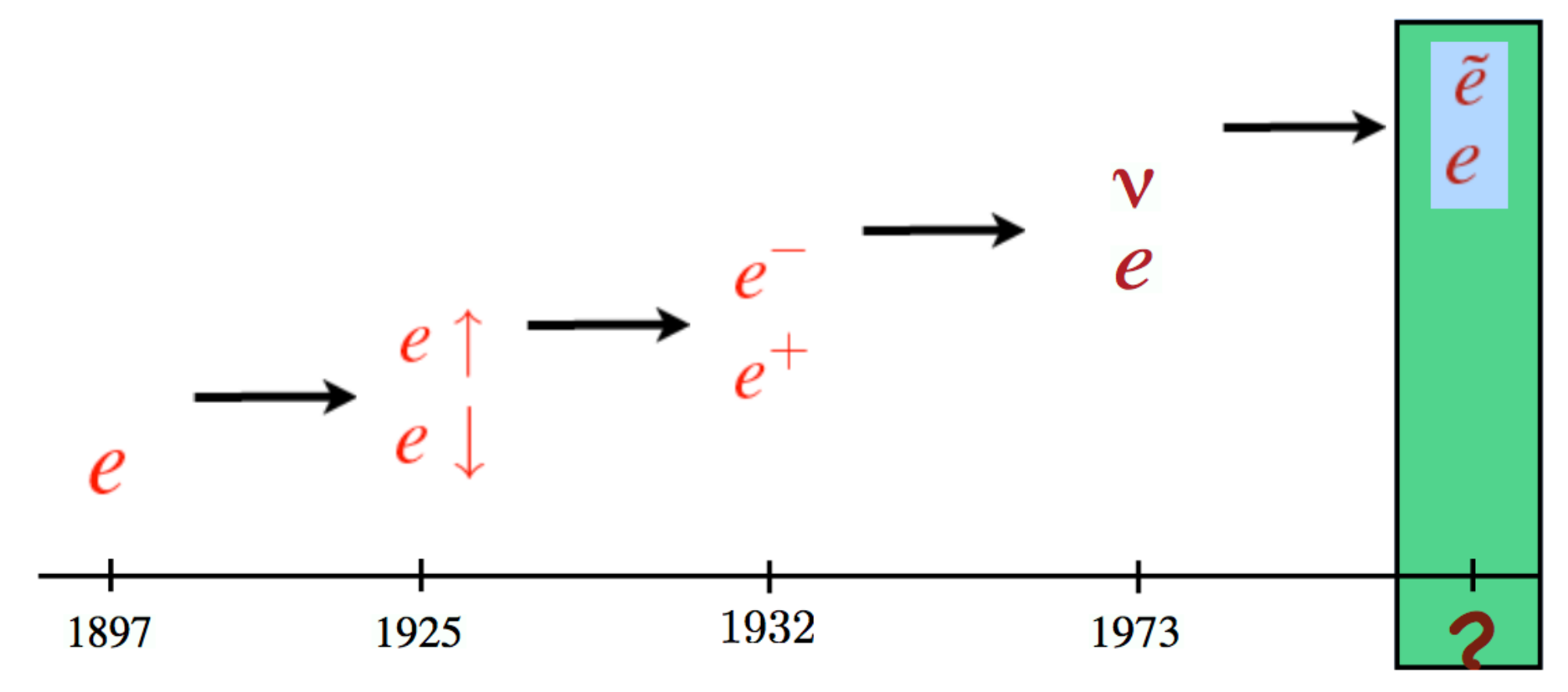}
\caption{A synthetic illustration of the role of symmetries in the history (and perhaps the future history) of the electron. }
  \label{history}
\end{figure}

Not the least drawback of the SM is that we cannot compute or not even estimate the Higgs mass, which is quadratically divergent. The classic way to address this issue has consisted in relating the Higgs mass to some other physical scale, not far from the Fermi scale itself and  not subject on its own to power-law divergences. From an aesthetic and general theoretic point of view supersymmetry  remains the relatively best solution of the hierarchy problem. All the history of PP seems to cry out for it, as synthetically but also effectively illustrated in Fig.~\ref{history}. 

The problem, not for supersymmetry in itself but for knowing if it is true in nature, is that the relation between the Higgs mass, or the Higgs vev, and the mass of the s-partners is a quantitative one only if one bars accidental cancellations among otherwise unrelated parameters. Although depending on peculiar configurations of the s-particle spectrum or on particular extensions of the Minimal Supersymmetric Standard Model, some level of accidental cancellations is now needed to be consistent with the LHC negative searches so far. This motivates the question: Is "low energy" supersymmetry still alive? The answer cannot be a sharp one. The judgement is suspended, although with pretty clear reasons of concern, in my view.

The absence, so far, of new physics signals at LHC raises in fact more general questions:
\begin{itemize}
\item  Is the quest for "naturalness" still relevant? 
\item How about: Naturalness = "low energy" New Physics?
\end{itemize}
The answer to the last question, I think, is clear. It is not a "theorem" anymore, as we used to say before or even at the time of LEP, because the experiments so far do not support it.  Are we lacking a clever "IR-UV connection", perhaps capable to solve as well  the cosmological constant  problem, without the need of "nearby" physics?  Maybe, but the difficulties  of decoupling the "low energy" scales from the Planck scale or from the Landau poles of non-asymptotically free couplings in the SM or from whatever "high energy" physical scale cold exist, remain there as a stumbling block. Lastly, does the multiverse emerge, supplemented by anthropic considerations,  as the "only solution" of the naturalness problem? A "solution" is such, in my view, only if it is supported by independent evidence, like the s-partners or the Higgs form factors in the case of the standard putative solutions. I do not yet see independent evidences in the case of the multiverse.  In a way or another, therefore, 
I think that the quest for naturalness remains relevant more than ever.

%

\section{Examples of interesting directions to be followed}
\label{sec:ewpt}

This is a time of healthy uncertainty in PP, which causes the emergence of new directions of research. Among the many possibilities in the second part of this discussion I want to describe briefly two such directions that look relevant and, most importantly, both have clear experimental implications.
\subsection{Minimal Mirror Twin Higgs}
\label{sec:MMTH}

The hypothesis of the existence of a Mirror Sector, an identical copy of the SM, has a long history\cite{Lee:1956qn,Kobzarev:1966qya}. Two key results follow from introducing an approximate spacetime parity symmetry, $P$, that exchanges the two sectors. First, the SM Higgs boson can be understood as a pseudo-Goldstone boson via the Twin Higgs mechanism\cite{Chacko:2005pe}, thus explaining the absence of BSM signals at LHC with a modest amount of fine-tuning. Second, and in my view even more importantly, Dark Matter (DM) may be mirror baryons with a density expected to be the same order as the baryon density, since both due to similar matter-antimatter asymmetries\cite{Goldberg:1986nk}.

For all this to work, $P$ must be broken, in particular to avoid inconsistencies with observations both in DM and in Dark Radiation (DR). Interestingly this can be done all at once by breaking $P$ in the Yukawa couplings\cite{Barbieri:2005ri,Barbieri:2016zxn}. Suppose in particular that the observed fermion mass hierarchy is described by suppression factors $\epsilon^{n_i}$ for charged fermion $i$, as can arise in Froggatt-Nielsen\cite{Froggatt:1978nt} and extra-dimensional theories of flavour\cite{Kaplan:2001ga}. The corresponding flavour factors in the mirror sector are $(\epsilon')^{n_i}$, so that spontaneous breaking of the parity $P$ arises from a single parameter $\epsilon'/\epsilon$. An overall  consistent picture emerges\cite{Barbieri:2017opf} for simple values of the $n_i$, that properly describe the observed fermion mass hierarchy.

\begin{figure}[t]
\centering
\includegraphics[clip,width=.45\textwidth]{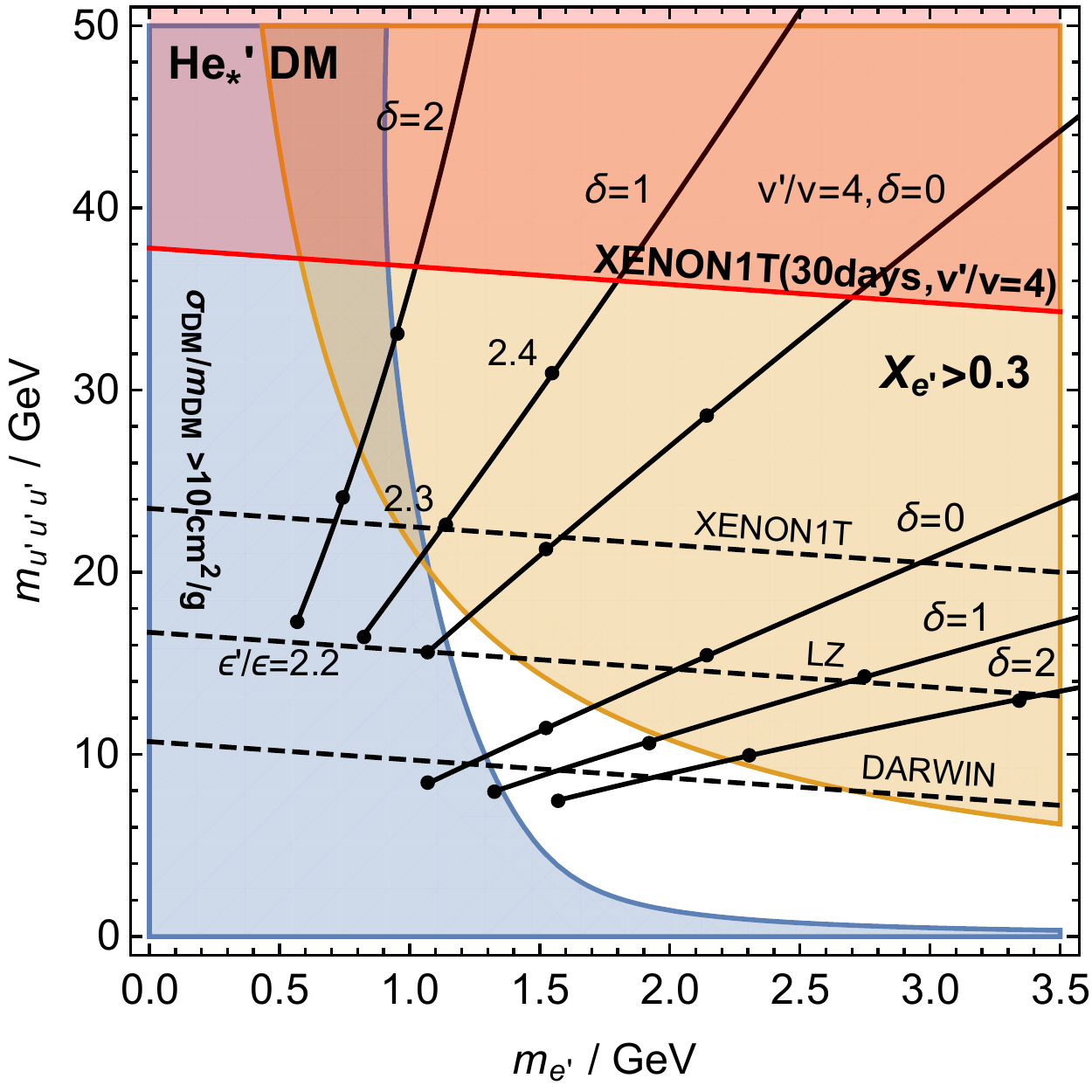}
\includegraphics[clip,width=.45\textwidth]{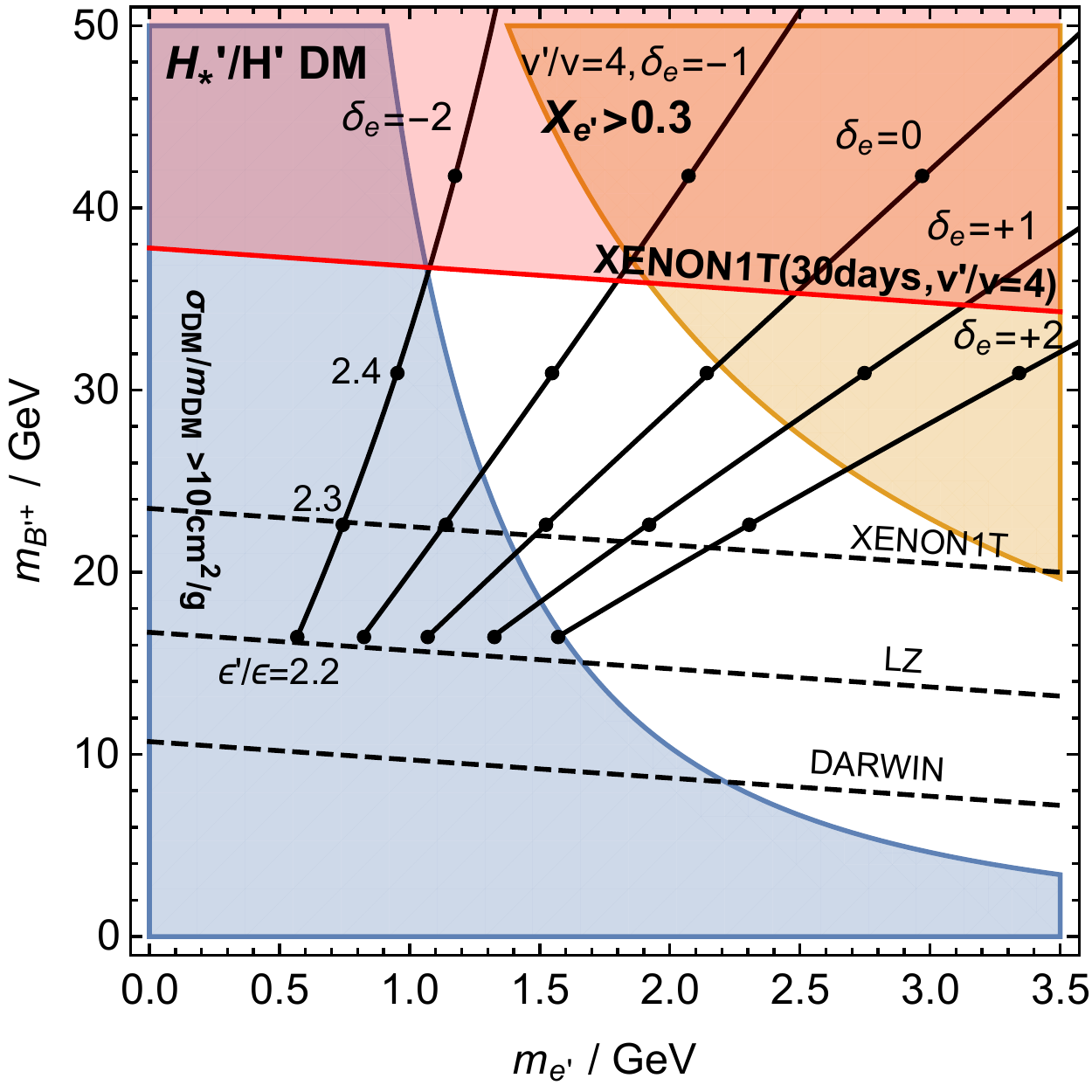}
\includegraphics[clip,width=.45\textwidth]{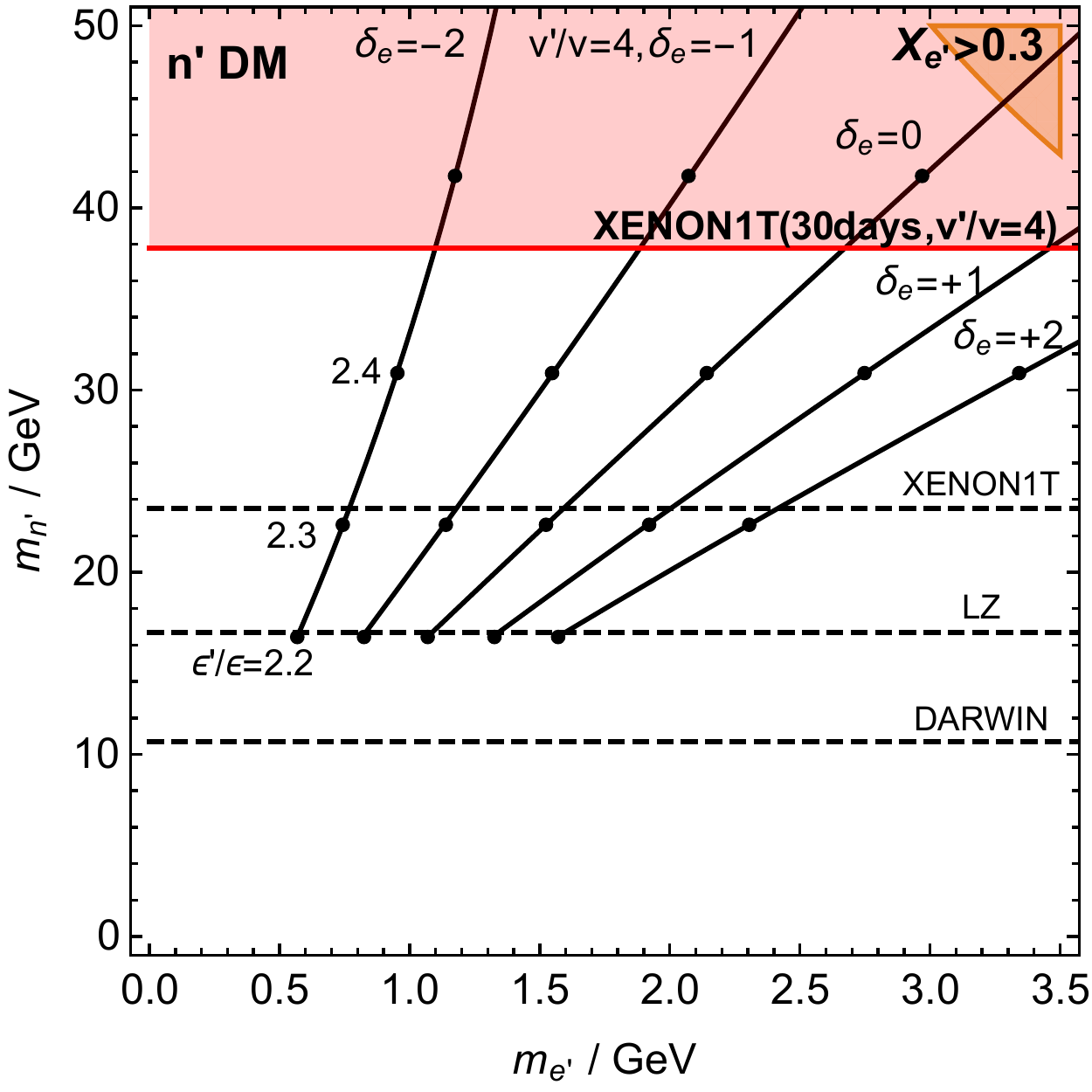}
\caption{
The masses of: i) charge-2 mirror baryon $B_{u'u'u'}$ (top left panel); ii) charge-1 mirror baryon $B^+=$$B_{d'd'd'}$/$B_{u'u'd'}$ (top right panel); iii) charge-0 mirror baryon $B_{u'd'd'}=n'$ (lower panel) shown as full lines against the mass of the mirror electron $e'$. Different  coloured areas give the excluded portions of the planes by 
 self-interactions of mirror atoms, by the mirror ionisation  fraction $X_{e'}$ and by direct detection\cite{Aprile:2017iyp}.
  Dashed curves give expected reaches of future direct detection experiments. From Ref.~\cite{Barbieri:2017opf}.
}
\label{fig:selfscatt-ioniz}
\end{figure}

The key experimental implications are:
\begin{itemize}
\item The existence of DR, in the form of mirror photons and mirror neutrinos, with a density between the $1\sigma$ and the $2\sigma$ upper bounds set by the current Planck data\cite{Ade:2015xua}.

\item A universal reduction of the Higgs signal strengths by 5 to 20 $\%$, due to a mixing between the original Higgs and mirror Higgs fields.

\item The existence of hydrogen-like or helium-like or neutron-like DM, depending on which one is the lightest, with properties broadly predicted in terms of $\epsilon'/\epsilon$ and of an $\mathcal{O}(1)$ parameter $\delta$, as summarised in Fig.~\ref{fig:selfscatt-ioniz} together with various astrophysical and cosmological constraints, taken from Ref.~\cite{Barbieri:2017opf}.
\end{itemize}

%

\subsection{Tantalising deviations from the SM in flavour physics}
\label{sec:LFV}

Starting from the 2012 HICHEP conference in Melbourne up to very recent results, numerous hints have emerged of Lepton Flavour Universality (LFU) violations observed in semi-leptonic 
B-decays, as recently summarised in Ref.~\cite{Buttazzo:2017ixm} and references therein. While not a single LFU ratio measurement exhibits a deviation with respect to the SM above the $3\sigma$ level, the overall set of observables is pretty much consistent and, once combined, the probability of a mere statistical fluctuation is very low. At the underlying quark level, the evidences collected so far can be grouped into:
\begin{itemize}
\item Deviations from $\tau/\mu$ (and $\tau/e)$ universality in $b\rightarrow c l\nu$ charged currents;
\item Deviations from $\mu/e$  universality in $b\rightarrow s l\bar{l}$ neutral  currents;
\end{itemize}
both at $\mathcal{O}(10\%)$ level relative to the corresponding SM amplitudes.
Caution in accepting evidence for LFU violation, not seen before in whatever process, is never enough. Yet the overall set of observables justifies the strong reaction of the theoretical community. 

The explanation of any putative deviation from the SM in flavour physics must be well designed enough to respect the  full body of experimental constraints in flavour physics, by now significantly extended. To this purpose  I advocate a weak form of Minimal Flavour Violation, as arising from a minimally broken $U(2)^n$ flavour symmetry, with the different $U(2)$ factors referring to the different irreducible representations of the gauge group. Under $U(2)^n$ the fermions of the third generation transform as singlets and the first two generations as doublets\cite{Barbieri:2011ci,Barbieri:2012uh}. 

$U(2)^n$ introduces a natural distinction between semi-leptonic $K$ and $\pi$ decays on one side, where tests of LFU are at the per-mil level, and $B$ semi-leptonic decays on the other side, where LFU violations seem to emerge. Furthermore $U(2)^n$ makes plausible that a massive bosonic mediator around the TeV scale, responsible for the new effect, be predominantly coupled to the third generation of fermions, with the couplings to the lighter generations only coming after small $U(2)^n$ breaking. Suppose that the mediator is a leptoquark, singlet under $U(2)^n$. Since leptons and quarks of the first two generations transform as doublets of different $U(2)$ factors, in the unbroken $U(2)^n$ limit such leptoquark indeed  couples to the third generation only. This hypothesis provides a natural first-order explanation for  the different size of the charged current versus the neutral current effects, about equally deviating from the  tree-level versus  loop-level SM amplitudes respectively: $b\rightarrow c \tau\nu$  only involves a single second generation particle, whereas $b\rightarrow s \mu\bar{\mu}$ has three light generation fermions\cite{Barbieri:2015yvd}. 

Already analysed from an EFT point of view\cite{Bordone:2017anc}, this broad picture leads to several testable predictions, among which:
\begin{itemize}
\item $b\rightarrow c(u) l\nu$: Universal deviation from unity of the Branching Ratios, normalized to the SM, for $B\rightarrow D^*\tau\nu, B\rightarrow D\tau\nu, \Lambda_b\rightarrow \Lambda_c\tau\nu,
B\rightarrow \pi\tau\nu, \Lambda_b\rightarrow p\tau\nu, B_u\rightarrow \tau\nu$;

\item $b\rightarrow s\tau\tau$: a large enhancement, relative to the SM, made plausible;

\item $b\rightarrow s\nu\nu$: $\mathcal{O}(1)$ deviations from the SM rate expected;

\item $K\rightarrow \pi \nu\nu$: $\mathcal{O}(1)$ deviations from the SM rate expected.

\end{itemize}

As already mentioned, the size of the effect in charged current calls for the existence of a massive mediator around the TeV scale, more likely in a strong coupling regime. From the theory side this will require a non trivial  UV completion of the phenomenological picture\cite{Buttazzo:2016kid,Barbieri:2016las,Megias:2017isd}. Experimentally a sizeable  (broad) excess in $pp\rightarrow \tau\tau$ and $pp\rightarrow bb,~tt$ is expected\cite{Buttazzo:2016kid,Faroughy:2016osc}.
\section{Conclusions}
\label{sec:conc}
Looking in part at the crystal  ball, I think that in the future of PP  a deeper theory then the SM is likely to emerge, of which theory the SM itself will in any case be a relevant limit. This comes from considering the incompletenesses of the SM, as opposed to  its great experimental successes: a kind of paradoxical situation, in my view. 

The very nature of fundamental physics and the current uncertain situation of PP  require and are already generating highly diverse directions of research. I have argued that precision measurements in Higgs and  flavour physics are to be vigorously pursued.  Imagine, at least as an example, what it would mean if the hints of LFU violation described  in Section \ref{sec:LFV} were confirmed. As I have already recalled, in the tradition of PP precision is a key to understanding. On the other hand many directions of research indicate that the boundaries between PP, cosmology and astrophysics are more and more fading away. The Mirror World briefly discussed in Section \ref{sec:MMTH} provides a motivated example.

\begin{acknowledgments}

I am grateful to Stefano Forte, Aharon Levy and Giovanni Ridolfi for taking the initiative of this book in memory and honour of Guido Altarelli. 
\end{acknowledgments}


\begingroup
\renewcommand{\addcontentsline}[3]{}
\renewcommand{\section}[2]{}

\endgroup
  
\end{document}